\documentclass[twocolumn,preprintnumbers,footinbib]{revtex4}
\usepackage{psfrag,epsfig,rotate,bm}
\begin{document}
\title{Microscopic theory of the inverse Faraday effect }
\author{Riccardo Hertel}
\email{r.hertel@fz-juelich.de}
\affiliation{Institute of Solid State Research (IFF), Research Center J\"ulich, D-52425 J\"ulich, Germany}  
\newcommand{\be}{\begin{equation}}
\newcommand{\ee}{\end{equation}}
\newcommand{\ber}{\begin{eqnarray}}
\newcommand{\eer}{\end{eqnarray}}
\newcommand{\p}{\partial}
\begin{abstract}
An analytic expression is given for the inverse Faraday effect, i.e. for the
magnetization occurring in a transparent medium exposed to a circularly
polarized high-frequency electromagnetic wave. Using a microscopic approach
the magnetization of the medium due to the inverse Faraday effect is
identified as the result of microscopic solenoidal currents generated
by the electromagnetic wave. In contrast to the better known
phenomenological derivation, the microscopic treatment provides
important information on the frequency dependence of the inverse
Faraday effect.

\end{abstract}
\maketitle
\section{introduction}
While the equations describing the transfer of energy and momentum
to a medium by an electromagnetic wave are commonly known from 
undergraduate students' texbooks, the study of the transfer of
angular momentum  of an electromagnetic wave has received relatively
little attention. The first prediction that circularly polarized light
should cause an angular momentum-flux is due to Sadovskii and dates
back to 1897. The original work of Sadovskii is written in Russian and
is difficult to find (see footnote 1 in Ref.~\cite{Beth36}). The first
detection of the Sadovskii effect, {\em i.e.} the mechanical measurement of a
torque exerted by circularly polarized light on a half-wave
plate, has been reported in 1936 by Beth \cite{Beth36} and Holbourn
\cite{Holbourn36}, independently.  The interesting history of the
study of angular momentum of electromagnetic waves is reported in 
several review papers \cite{Vulfson87,Sokolov91}. Since magnetic
moments have an angular momentum, 
the possibility to influence the magnetization in a magnetically ordered
material by means of circularly polarized light appears to be a straightforward 
continuation of the old experiments by Beth and Holbourn. It is
therefore remarkable that only very recently, about 70 years after these
historic experiments, it has been demonstrated by Kimel
{\em et al.} \cite{Kimel05} that spins can indeed be manipulated with
a circularly polarized laser beam. Apart from their importance for
fundamental processes concerning the magnetization dynamics, these
spectacular experiments are also interesting, because of their
potential technical implications. These studies may pave the way
towards an ultrafast, focussed laser-controlled magnetic writing
process that could eventually replace the comparatively clumsy data
storage technique being in use nowadays, which involves microscopic
coils and write heads that may generate undesired fringing fields. The 
effect reported by Kimel {\em et al.} is attributed to the generation of a
stationary magnetic field resulting from irradiation with circularly
polarized light, which is known as the inverse Faraday effect. 

The inverse Faraday effect (IFE) has first been predicted by Pitaevskii more
than 40 years ago \cite{Pitaevskii61}, while the name (IFE) seems to have 
been coined only four years later by van der Ziel {\em et al.}
\cite{vdZiel65,Pershan66}. The IFE was originally derived on a 
phenomenological basis and has received a certain publicity
among physicists, because it was briefly treated in some older
editions of a well-known textbook series on theoretical physics
\cite{Landau84}. A much more detailed description of the IFE has been
elaborated in the late seventies and the early eighties 
by theoretical plasma physicists, mainly in the former Soviet Union 
\cite{Tsytovich78,Belkov79,Kono81,Karpman82}. This literature is presumably
hardly known in the part of the scientific community that is currently
investigating ultrafast magnetization processes. The scope of this
paper is to provide a simple derivation of the IFE that involves
microscopic currents, described in more detail by Karpman {\em et
  al.}~\cite{Karpman82}, rather than a purely phenomenological
approach. In the wake of the recent studies by Kimel {\em et al.},
this short review of the theory of the IFE is likely to be of interest
for researchers working in the domain of magnetism who may not happen 
to be experts in theoretical plasma physics.

\section{Basic equations}
When a material is irradiated with a high-frequency (HF) electromagnetic
wave, the primary coupling between the wave and the material
is given by the interaction between the electrons and the electric field of
the wave. One may therefore neglect in a first approximation the
wave's fluctuating magnetic field $\delta\bm{B}$.

For simplicity, we also neglect any quantum mechanical effects and
assume that the band structure of the material is unimportant for the
motion of the electrons. In the case of a metallic material
the conducting electrons can be treated as a collisionless plasma in
which the electrons can move freely, at least on the time scale given
by the period of the HF field. Such an electron plasma can
conveniently be represented as a fluid with density $n(\bm{r},t)$
and velocity  $\bm{v}(\bm{r},t)$. 
The electron density $n$ and the velocity $\bm{v}(\bm{r},t)$ are related to
each other by the continuity equation 
\be\label{conti}
\frac{\p}{\p t}n+\bm{\nabla}\left(n\bm{v}\right)=0\quad.
\ee
The electron density $n$ and the velocity $\bm{v}$ give rise to
 an electric current density $\bm{j}$ according to 
\be
\bm{j}=en\bm{v}\quad, 
\ee
which, in the simplest approximation, is proportional to
an electric field $\bm{E}$
\be
\bm{j}=\sigma\bm{E}.
\ee
In the case of an oscillating electric field 
\be\label{wave}
\delta\bm{E}(\bm{x},t)=\bm{\hat{E}}\exp(i\bm{kr}-i\omega t)
\ee
of high frequency $\omega$ and wave vector $\bm{k}$, the conductivity
$\sigma$ of an isotropic, collisionless plasma is \cite{Landau84}  
\be\label{hfcond}
\sigma=\frac{i\langle n\rangle e^2}{m\omega}\quad,
\ee
where $m$ is the electron mass. The brackets $\langle\,\rangle$ denote
a time average of the electron density $n$ over several periods of the
HF field. 

For the following, only the temporal oscillation of the local field at
a given point $\bm{r}$  will be required,
%\be\label{wavet}
%\delta\bm{E}(t)=\bm{\hat{E}}\exp(-i\omega t)\quad,
%\ee
so that the spatial dependence of the
electromagnetic wave $\delta\bm{E}$ can be omitted. Note that the
amplitude $\hat{\bm{E}}$ is in general a complex quantity. The
physically relevant real value of the field is given by 
\be 
\Re\left\{\delta\bm{E}\right\}=\frac{1}{2}\hat{\bm{E}}\exp(-i\omega t) +c.c.\quad,
\ee
where $c.c.$ denotes complex conjugation of the previous term.

\section{separation of time scales}
To investigate the influence of an HF field on a plasma, it is useful
to treat different time scales separately. 
%The principle is quite similar to the Born-Oppenheimer approximation.
The time scales are split by representing the electron density $n$,
the velocity $\bm{v}$, and any other quantity $a$ which is required for
the description of the plasma as a sum of two parts  
\be
a=\langle a\rangle +\delta a\quad,
\ee
one of which ($\langle a\rangle$) is the time-averaged value that is 
constant or slowly changing on the time scale given by the period
of the HF field, and the other ($\delta a$) is an oscillating part
with the same time dependence $\propto\exp(-i\omega t)$ as the HF
field \cite{Gorbunov73}. 

We may safely assume that the magnitude of the oscillating part
$\delta n$ of the electron density, i.e. the size of the density
fluctuations induced by the HF field, is much smaller than the  
stationary value of $n$: 
\be\label{smaller}
\delta n\ll \langle n \rangle\quad.
\ee
Inserting the separation of stationary and oscillatory quantities
into the continuity equation (\ref{conti}) yields
\be\label{conti_pert}
\frac{\p}{\p t}\left(\langle n\rangle+\delta n\right)+\bm{\nabla}\left(\langle n\rangle
\langle \bm{v}\rangle +\langle n\rangle \delta\bm{v}+
\delta n \langle \bm{v}\rangle+\delta n \delta\bm{v}
\right)=0.
\ee
Owing to $\langle\delta a\rangle=0$, the time-averaged continuity
equation is 
\be\label{conti_sts}
\frac{\p}{\p t}\langle n\rangle+\bm{\nabla}\left(\langle n\rangle
\langle \bm{v}\rangle +\langle\delta n
\delta\bm{v}\right\rangle)=0\quad. 
\ee
This is the so-called slow time scale of the continuity equation. 
Using the approximation of eq.~(\ref{smaller}), the fast time scale reads
%\be\label{conti_fts}
%\frac{\p}{\p t}\delta n+\bm{\nabla}\left(\langle n\rangle
%\left(\delta\bm{v} +\frac{\delta n}{\langle n\rangle}  \langle\bm{v}\rangle
%\right)\right)=0\quad,
%\ee
%as can be seen by subtracting (\ref{conti_sts}) from (\ref{conti}). 
%The approximation (\ref{smaller}) yields
\be\label{fastn}
\frac{\p}{\p t}\delta n+\bm{\nabla}\left(\langle n\rangle
\delta\bm{v}\right)=0\quad.
\ee
%Here, the approximation of eq.~(\ref{smaller}) has been used.
%which also results if we assume that there is no stationary current in
%the plasma, {\em i.e.} $\langle\bm{v}\rangle=0$.
The same approximation yields the oscillating part $\delta\bm{j}$
of the current density $\bm{j}$ as 
\be\label{fastj}
\delta\bm{j}=e\langle n\rangle\delta\bm{v}\quad.
\ee
Hence, the fluctuating part of the velocity is
\be\label{fastv}
\delta \bm{v}=\frac{\sigma}{\langle n\rangle e}\delta\bm{E}\quad.
\ee
According to eq.~(\ref{fastn}) and eq.~(\ref{fastv}), the fluctuating
part of the electron density $n$ is given by
\be
\delta n 
%& = & -\frac{i}{\omega e}\bm{\nabla}\delta\bm{j}\nonumber \\
 = -\frac{i}{\omega e}\bm{\nabla}\left(\sigma\delta\bm{E}\right)\quad.
\ee
The slow time scale term of the current density $\langle
\bm{j}\rangle$ is  
\be
\langle\bm{j}\rangle=e\left(\langle n\rangle\langle\bm{v}\rangle
+\langle\delta n\delta\bm{v}\rangle\right)
\ee
or simply 
\be\label{slowj}
\langle\bm{j}\rangle=e\langle\delta n\delta\bm{v}\rangle
\ee
if $\langle\bm{v}\rangle=0$, which shall be assumed henceforth.

\section{inverse Faraday effect}
Equation (\ref{slowj}) describes the occurrence of a stationary
current density $\langle\bm{j}\rangle$ that is solely due to the  
high-frequency oscillations of the electron density and velocity
induced by the HF field. This term will eventually lead to an
expression for the IFE, {\em i.e.} for the magnetization of the plasma
in the field of a circularly polarized electromagnetic wave.

To further evaluate the right-hand side of eq.~(\ref{slowj}) it
should be beared in mind that the term $\delta n\cdot\delta\bm{j}$ 
describes the product of two {\em real} quantities.
Therefore,
\ber
\delta n\cdot\delta\bm{v} & =& \Re\left\{ \hat{n}\exp(-i\omega t)\right\}\cdot
\Re\left\{\hat{\bm{j}}\exp(-i\omega t)\right\}\nonumber \\
%&=&\frac{1}{2}\left\{\hat{n}\exp(-i\omega t)+\hat{n}^*\exp(i\omega
%  t)\right\}\nonumber\\
%&&\times\frac{1}{2}\left\{\bm{\hat{j}}\exp(-i\omega t)+\bm{\hat{v}^*}\exp(i\omega
%  t)\right\}\cdot\nonumber\\
&=&\frac{1}{4}\left[  \hat{n}\bm{\hat{v}}^*
+\hat{n}^*\bm{\hat{v}}+\hat{n}\bm{\hat{j}}\exp(-2i\omega
t)+\hat{n}^*\bm{\hat{v}}^*\exp(2i\omega t)
%\right.\nonumber\\ &&\left.
\right]
\nonumber\\
\eer
and hence
\ber
\langle\delta n\delta\bm{v}\rangle
%&=&\frac{1}{2}\Re \left\{ \hat{n} \hat{\bm{v}} \right\}\nonumber \\
&=&\frac{1}{4} \left(\hat{n} \hat{\bm{v}}^*+\hat{n}^*
\hat{\bm{v}}\right) \nonumber \\  
&=&-\frac{i}{4e^2\langle n\rangle\omega}
\left[\sigma^*\hat{\bm{E}}^*\cdot\bm{\nabla}\left(\sigma\hat{\bm{E}}\right)-{\rm
    c.c.}
%\sigma\hat{\bm{E}}\cdot\bm{\nabla}\left(\sigma^*\hat{\bm{E}}^*\right)
\right].\nonumber\\
\eer
Using the identity 
\be
\bm{A}\cdot\bm{\nabla}B-\bm{B}\cdot\bm{\nabla}A=
\bm{\nabla}\times\left(\bm{A}\times\bm{B}\right)-\left(\bm{B\nabla}\right)\bm{A}
+\left(\bm{A\nabla}\right)\bm{B}
\ee
one obtains 
\be\label{jwithg}
\bm{\langle\bm{j}\rangle}=-\frac{i}{4e\langle n\rangle\omega}
\bm{\nabla}\times\left[\sigma^*\hat{\bm{E}}^*\times\sigma\hat{\bm{E}}\right]+
\bm{\Gamma}  
\ee
with 
\be
\bm{\Gamma}=\frac{1}{4e\langle
  n\rangle\omega}\left[\left(i\sigma^*\hat{\bm{E}}\bm{\nabla}\right)
  \left(\sigma\hat{\bm{E}}\right)+{\rm c.c.}\right] \quad.    
\ee
The term $\bm{\Gamma}$ describes the so-called ponderomotive force acting
in the plasma. The ponderomotive force leads to a special type of 
stationary currents generated by a HF field. These currents 
result from spatial inhomogeneities of the plasma or of the field of
the wave. This term $\bm{\Gamma}$ is discussed in detail, {\em e.g.}, in
Ref.~\cite{Karpman82}. Although it can be assumed that ponderomotive
forces also play a role in the recent experiments by Kimel {\em et al.}, 
this term does not describe the IFE. 
It is the first term on the right-hand side of eq.~(\ref{jwithg}) that
has the form of a solenoidal magnetization current $\bm{j}_{\rm m}$:
\be\label{crotm}
\bm{j}_{\rm m}=c\bm{\nabla}\times\bm{M}\quad.
\ee
Hence, the magnetization $\bm{M}$ generated in the plasma by the HF field is
\ber
\bm{M}&=&-\frac{im\omega_{\rm p}^2}{16\pi e^3c\langle  n\rangle^2\omega}
\left[\sigma^*\hat{\bm{E}}^*\times\sigma\hat{\bm{E}}\right] \\
%\ee
%\ber
%\bm{M}
&=&\frac{ie\omega_{\rm p}^2}{16\pi\omega^3mc}
\left[\hat{\bm{E}}\times\hat{\bm{E}}^*\right]\label{inverse} 
\eer
where $\omega_{\rm p}=(4\pi\langle n\rangle e^2/m)^{1/2}$ is the plasma frequency.
%\ee
%

\section{discussion}
It should be noted that eq.~(\ref{inverse}) has been derived
without making any assumption concerning the polarization of the
electromagnetic wave. 
In the case of a circularly polarized wave propagating in the $z$
direction, the last term on the right hand side can be written as
\be\label{circular}
\hat{\bm{E}}\times\hat{\bm{E}}^*=\pm i\left|\bm{E}\right|^2\cdot\bm{e}_z\quad,
\ee
while it is equal to zero in the case of linear polarization.
The plus and the minus sign in eq.~(\ref{circular}) refers to left and 
right circular  polarization, respectively. Therefore, $\bm{M}$ is a
real-valued, stationary magnetization that is induced in the medium by
a circularly polarized electromagnetic wave. The magnetization is
parallel to the axis of propagation of the wave and its sign depends
on the chirality of the wave.

If the IFE is applied to change the magnetization in
magnetically ordered materials, the magnetization induced by the HF
field according to eq.~(\ref{inverse}) should reorient
the magnetic moments in the sample, leading to a permanent 
magnetization that persists after the HF field is switched off. 
The mechanism that converts the field-induced magnetization currents into
a  permanent magnetization is not yet clear.  The simplest
explanation would be the magnetostatic alignment of the magnetic
moments in the field generated by the IFE. However, a more
complicated mechanism on an electronic level cannot be ruled out
currently. 

In the derivation of the IFE according to eq.~(\ref{inverse}) it has
been assumed that the electromagnetic wave is not absorbed by the
medium. Since the value of the  high-frequency conductivity is purely imaginary,
cf.~eq.~(\ref{hfcond}), the induced current density
$\delta\bm{j}$ is phase-shifted by $\pi/2$ with respect to the
electromagnetic wave.
%\be
%\bm{j}\exp(-i\omega t)\propto i\bm{E}\exp(-i\omega
%t)=\bm{E}\exp(-i\omega t+i\pi/2)\quad.
%\ee
Hence, the power density transferred to the plasma
\be
\langle\delta\bm{j}\delta\bm{E}\rangle
%\frac{ie^2n}{m\omega}\langle\delta\bm{E}\delta\bm{E}\rangle
=\frac{ie^2n}{2m\omega}\left|\bm{E}\right|^2\quad
\ee
is also purely imaginary, which means that the electromagnetic wave is not absorbed.
However, it should not be concluded from this absorption-free
derivation of the IFE that the absorption of a
circularly polarized electromagnetic wave
does {\em not} lead to a magnetization of the medium. 

Interestingly, none of the historical arguments on the transfer of angular
momentum mentioned in the introduction has been used for the
derivation of the IFE. In fact, the investigation of the angular
momentum transported by a circularly polarized electromagnetic plane
wave leads to several complications. It has been demonstrated, that an
infinitely extended, circularly polarized plane wave, surprisingly,
does not carry any angular momentum \cite{Sokolov91}. This
counterintuitive result is not in contradiction to the experimental
proof of the angular momentum demonstrated by Beth \cite{Beth36} and
Holbourn \cite{Holbourn36}, since it has also been shown that a real,
collimated beam of circularly polarized light {\em does} have an
angular momentum flux \cite{Vulfson87}. Using higher order terms in the
so-called geometric optics approximation \cite{Kravtsov69}, an analytic
expression for the spin flux density of a collimated circularly
polarized wave in a plasma has  been reported \cite{Kirochkin93}.  
However, angular momentum considerations are not necessary and don't
seem to be helpful for the derivation of the IFE.

\section{conclusion}
The ongoing research efforts to find faster and more precise methods
to store information in increasingly miniaturized magnetic 
particles has lead to several successful collaborations between scientists
specialized in different domains. This is especially true for 
modern experimental investigations, {\em e.g.}, on ultrafast
magnetization process in nanostructures, where usually several
different techniques are involved. The recently demonstrated
manipulation  of spins by means of a circularly polarized laser beam is a further
advance in this field, which requires an expansion of the knowledge
about magnetization processes driven by electromagnetic waves. Similar
to the positive synergy effects that have been obtained in
experiments, the theory can also benefit significantly by combining
the theory of magnetism with theoretical plasma physics, in which the
interaction of electrons with radiation is of central importance.  

Compared to the relatively well known phenomenological treatment of
the IFE, the microscopic derivation of the IFE presented in this paper
gives a clearer insight into the processes leading to the
magnetization of the sample. A further important aspect is that
eq.~(\ref{inverse}) is free of material constants and thus allows for
quantitave predictions of the IFE. In particular, a strong dependence
of the magnetization generated by the IFE on the frequency of the
applied field $M\propto\omega^{-3}$ is predicted. An experimental
investigation on the frequency dependence of the IFE would be 
desirable to validate this equation.

\section*{Acknowledgements}
I would like to thank Prof.~U.~Schumacher and Dr.~E.~R\"auchle for
introducing me into this subject several years ago. I also want to
thank Prof.~C.M.~Schneider for reading and commenting on the
manuscript.

%%%%%%%%%%%%%%%%%%%% References %%%%%%%%%%%%%%%%%%%%%%%%%%%%%%
\bibliography{/tmp_mnt/users/iff_ee/hertel/bib/plasmabib.bib}
\bibliographystyle{phreport}
\end{document}